\documentclass{icrc}

\usepackage{times}
 \usepackage{graphicx} 

\begin{document}

\title{A multivariate study of mass composition for simulated showers
 at the Auger South Observatory.}
\author[1]{G. A. Medina-Tanco}
\author[2]{S.J.Sciutto}
\affil[1]{Instituto Astronomico e Geofisico, Universidade de S\~ao
Paulo , Brazil}
\affil[2]{Departamento de F\'{\i}sica, Universidad Nacional de La
Plata, C. C. 67, 1900 La Plata, Argentina}

\correspondence{Medina-Tanco (gustavo@iagusp.usp.br)}

\firstpage{1}
\pubyear{2001}


\maketitle

\begin{abstract}
The output parameters from the ground array of the Auger South
observatory, were simulated for the typical instrumental and
environmental conditions at its Malarg\"ue site using the code
sample-sim. Extensive air showers started by photons, protons and
iron nuclei at the top of the atmosphere were used as triggers.
The study utilized the air shower simulation code Aires with both
QGSJet and Sibyll hadronic interaction models. A total of 1850
showers were used to produce more than 35,000 different ground
events. We report here on the results of a multivariate analysis
approach to the development of new primary composition
diagnostics.
\end{abstract}

\section{Introduction}

The experimental detection of ultra high energy cosmic rays
($E>10^{20}$ eV) poses some of the most exciting problems in
modern astrophysics. Up to now no astrophysical objects are known
that could accelerate charged particles to such energies. If the
sources are located on cosmological distances, then it would be
expected that the Cosmic rays arriving to the Earth will loose
energy after interacting with the cosmic microwave background,
until reaching a threshold energy of about $6\times 10^{19}$ eV.
This energy would therefore mark a sharp end of the Cosmic Ray
spectrum. No such sharp end is seen by experiment so far. If the
sources are nearby, then an anisotropic distribution of arrival
directions is expected because in this case the directions of
arrival would point to the sources.

Alternative explanations of the existence of the Ultra high energy
Cosmic Rays have been developed by theorists over the last few
years: New particles, new physics or exotic phenomena, such as
decaying topological defects, or the violation of Lorentz
invariance.

To effectively check any of these "classic" or alternative
theories it is necessary to measure with adequate statistics the
highest energy Cosmic Rays. It is necessary to accurately
determine the form of the spectrum, the distribution of arrival
directions over the whole sky, and the identity of the particles.

The Auger Observatory \citep{auger} has the aim of collecting
enough experimental data to give appropriate answers to those
questions. It consists in two detectors of 3000 km$^2$ each,
positioned on the Southern and Northern hemispheres. Each detector
will be capable of measuring the properties of the showers
generated by the ultra high energy cosmic rays. An array of
surface detectors (SD) will measure the characteristics of the
shower particles reaching ground level, while a fluorescence
detector will measure the light emitted after the interaction of
the shower particles with the atmosphere.

The development of extensive air showers (EAS), as characterized
by lateral distribution, curvature of the shock front, rising
time, pulse shape, total number of photoelectrons, etc., carry
information regarding the direction, energy and identity of the
incoming primary. However, while direction and energy can be
estimated rather easily from ground array data (e.g.
\citet{Billoir2000}), the definition of a convenient and efficient
diagnostic for primary identity discrimination remains a
challenging issue.

In particular, besides some punctual indications against UHE
photons as primaries \cite{Bird95,Halzen95,Nagano99}, only one
comprehensive study limiting the photon flux above $10^{19}$ eV
has been published \cite{Ave00} up to now, and it is based on an
analysis of inclined showers at Haverah Park (zenith angles $>
60^{o}$). The separation between light (protons) and heavier (Fe
nuclei) hadrons is still much more difficult.

In this paper we present preliminary results of an ongoing effort
to develop primary identification diagnostics with the aid of
multivariate techniques. A pragmatic approach is taken to the
practical problem of statistically determining the identity of the
primaries starting EAS at the top of the atmosphere with the
ground array of the Auger observatory as the specific target.

\section{Principal component analysis: photon-hadron separation}

A large sample of showers for primary photons, protons and iron
nuclei is generated with the AIRES code and, transformed into
ground array events of a model Auger observatory, used to trigger
the surface detectors, simulated with the sample-sim code.

The AIRES system is a set of programs to produce simulations of
air showers, and to analyze the corresponding data. All the
relevant particles and interactions are taken into account during
the simulations, and a number of observables are measured and
recorded, among them, the longitudinal and lateral profiles of the
showers, the arrival time distributions, and detailed lists of
particles reaching ground that can be further processed by
detector simulation programs. The AIRES system is explained in
detail elsewhere \citep{aires1,aires0}.

The showers processed in this work were generated with the AIRES
system, and consist in a series of 1831 proton, gamma, and iron
showers, with energies in the range $10^{17.5}$ eV to $10^{20.5}$
eV, and zenith angles in the range 0 to 60 degrees. Each shower is
reused 20 times at diferent location in the array, and so the
final number of available events is $36620$. The hadronic models
used are QGSJET \citep{QGSJET} and Sibyll \citep{SIBYLL}.

The surface detectors have been simulated using the "sample-sim"
SD simulation program \citep{Billoir2000}.

The directly observable output for each event, which include the
number and spatial distribution of triggered tanks and the time
profile of the signal at each station, together with more easily
reconstructed quantities (e.g., energy and zenith angle) are used
to define different sets of parameters. Each set of parameters
constitutes an n-dimensional orthogonal space which is later
studied using principal component analysis (PCA) in search for
primary separation.

The PCA method simply performs a rotation in the n- dimensional
space to a new orthogonal coordinate system whose unit vectors are
the eigenvectors of the system. These new axis have a special
meaning, since their associated eigenvalues are a measure of the
dispersion of the data along each axis. Thus, the principal
eigenvector has the largest associated eigenvalue, and therefore
the largest dispersion, or information content, of the sample; the
second eigenvector has the second largest dispersion and so on.
Typically, one can quantify the amount of information associated
with a subset of axis, and can even expect to uncover the true
dimensionality of the system if this has been overestimated.

One advantage of the PCA method is that, involving only rotations,
the new axis are only linear combinations of the original
magnitudes.

As an illustrative example, lets take a parameter space defined
arbitrarily by:

\noindent a (sort of) curvature estimator,

\begin{equation}
P_{1}= \left[ \frac{\langle T_{0,ext} \rangle - \langle T_{0,int}
\rangle}{\langle r_{ext} \rangle - \langle r_{int} \rangle}
\right] \times \sin{\theta}
\end{equation}

\noindent where the subscripts "ext" and "int" refer to stations
that are farther away and nearer the shower axis than the median
distance $r_{c}$ of the triggered stations, and $r_{ext}$ and
$r_{int}$ are the average distances inside each region.

\noindent the third largest total number vertical equivalent
muons, $N_{vem,i}$,

\begin{equation}
P_{2}= \left[ N_{vem,i} \right]_{3rd}
\end{equation}

\noindent pulse shape/rising time (average),

\begin{equation}
P_{3} = \langle \frac{T_{50}}{T_{10}+T_{50}} \rangle
\end{equation}

\noindent where $T_{i}$ are the fluence times for 10\% and 50\% of
the total fluence at a given station.

\noindent pulse shape/rising time (3rd largest value),

\begin{equation}
P_{4} = \left( T_{10}+T_{50}+T_{90} \right)_{3rd}
\end{equation}

\noindent (sort of) lateral distribution,

\begin{equation}
P_{5} = \left[ \frac{N_{vem,i}}{P_{4}} \right]_{5th} /
        \left[ \frac{N_{vem,i}}{P_{4}} \right]_{3rd}
\end{equation}

\noindent rising time (3rd largest value),

\begin{equation}
P_{6}=\left( \frac{T_{10}-T_{0}}{T_{90}-T_{0}}  \right)_{3rd}
\end{equation}

\noindent plus: the median of the station distances to the axis of
the shower, $P_{7}=r_{c}$, primary energy, $P_{8}=E$, zenith
angle, $P_{9}=\theta$, number of triggered stations,
$P_{10}=N_{stat}$. All these parameters are later normalized so
that their dynamical ranges are in the interval $(-1,1)$.

When a PCA analysis is performed in this parameter space, it is
found that the first 4 eigenvectors are responsible for $\sim
80$\% of the variance (or information content) of the system. The
7th eigenvector is responsible for only $\sim 6$ \% of the
variance.

The best separation between nuclei and photons is obtained for the
projection onto the plane defined by the first and seventh
eigenvectors (see figure 1). The thick line, $EV_{7} = -48.89
\times (EV_{1}+0.007)^{2}+0.011$, leaves only 0.8\% of the nuclei
in the region of photons and 12\% of the photons in the region
corresponding to nuclei. Therefore, the probability of
misidentifying a photon is $2.7$\% and the probability
misidentifying a nuclei is $3.8$\%.

\begin{figure}[t]
\vspace*{2.0mm} 
\includegraphics[width=8.3cm]{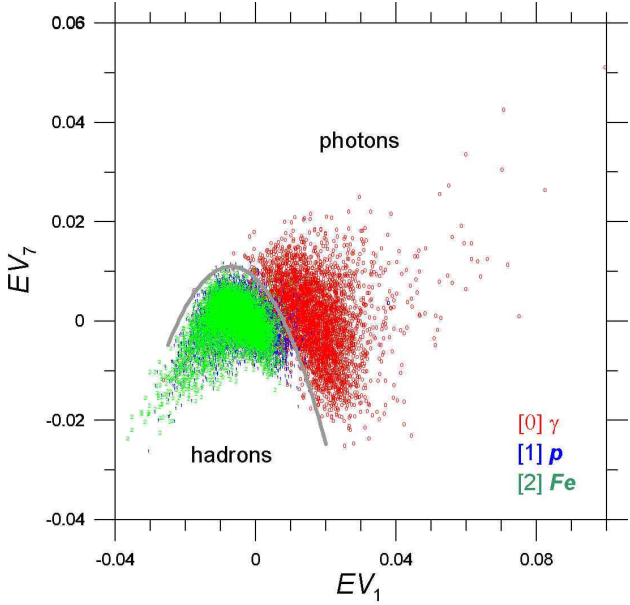} 
\caption{PCA results on the illustrative parameter space. The best
separation between nuclei and photons is obtained for the
projection on the plane defined by the first and seventh
eigenvectors. The thick line misclassifies $3.8$\% of the nuclei
as photons and $2.7$\% of the photons as nuclei.}
\end{figure}

Once the photons have been separated, the same process can be
applied to nuclei alone. However, as was stated before, this is a
much more complicated problem as shown in figure 2. The
optimization of of a diagnostic method in this case is still
ongoing work.

\begin{figure}[t]
\vspace*{2.0mm} 
\includegraphics[width=8.3cm]{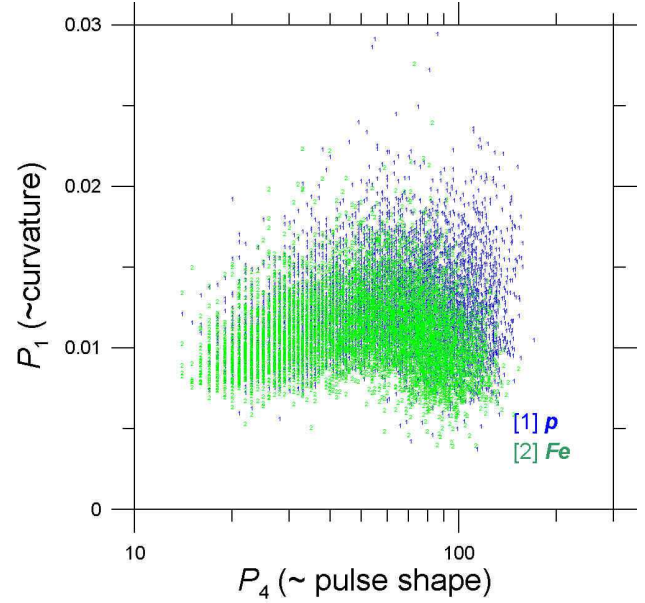} 
\caption{Projection onto the $P_{1}$--$P_{4}$ plane of the sample
points, once the photon events have been extracted, showing the
difficulty involved in the separation of light and heavy nuclei.}
\end{figure}

\section{Neural Network approach for p-Fe separation}

\subsection{QGSJet hadronic interaction model}

An alternative approach for hadronic primary separation can be
obtained by applying neural network technics to the problem.

An artificial neural network consists of a set of simple
processing units which communicate by sending signals to each
other over a large number of weighted connections. In general
terms, neurons are structured in an array of hidden layers bounded
by input and output slabs. Each unit receives inputs from
neighbors or external sources and computes an output, $y_{k}$,
which is propagated to other units:

\begin{equation}
y_{k} = F_{k} \left( \Sigma_{j} w_{jk} \times y_{j} + b_{k}
\right)
\end{equation}

\noindent where the sum extends over all the units $j$ effectively
connected to $k$,  $y_{j}$ is the input to unit $k$ coming from
unit $j$, $w_{jk}$ is the corresponding weight for that connection
and $b_{k}$ is a bias or offset term. $F_{k}$ is the {\it transfer
function\/}, usually a nondecreasing function of the total input.
Weights are the result of a training process in which known
input-output pairs are fed to the network.

As an example of this powerful method, in figure 3 we show the
results for a feed forward network, i.e., data flows exclusively
from input to output -- no feedback present
\citep{feedforward_Rumelhart,feedforward_Hagan,neuralnetworks_Krose},
constituted by four layers of neurons with 3, 20, 3 and 1 neurons
respectively, with tan-sigmoid (hidden) and log-sigmoid (output)
transfer functions. The network was trained using the resilient
backpropagation training algorithm in order to overcome problems
arising from the small derivative of the sigmoid function far from
the origin.

The input parameters used, based on direct observables and
reconstructed magnitudes from the surface array detector, are:

\begin{equation}
P_{1} = \frac{1}{N_{stat}} \times
        \Sigma_{i=1}^{N_{stat}}
        N_{vem,i} \left(  \frac{r_{0,i}}{1000 m} \right)^{3}
\end{equation}

\begin{equation}
P_{2} = \frac{1}{N_{stat}} \times
        \Sigma_{i=1}^{N_{stat}}
        \left( T_{0,i} - T_{sp,i} \right)
        \left(  \frac{r_{0,i}}{1000 m} \right)^{-2}
\end{equation}

\begin{equation}
P_{3} = \frac{1}{N_{stat}} \times
        \Sigma_{i=1}^{N_{stat}}
        \left( T_{10,i} - T_{0,i} \right)
        \left(  \frac{r_{0,i}}{1000 m} \right)^{-1}
\end{equation}

\begin{equation}
P_{4} = \frac{1}{N_{stat}} \times
        \Sigma_{i=1}^{N_{stat}}
        \left( T_{50,i} - T_{0,i} \right)
        \left(  \frac{r_{0,i}}{1000 m} \right)^{-1}
\end{equation}

\begin{equation}
P_{5} = \frac{1}{N_{stat}} \times
        \Sigma_{i=1}^{N_{stat}}
        \left( T_{90,i} - T_{0,i} \right)
        \left(  \frac{r_{0,i}}{1000 m} \right)^{-1}
\end{equation}

\noindent plus energy ($P_{6}$), zenith angle ($P_{7}$) and number
of triggered stations  ($P_{8}$); where $N_{stat}$ is the number
of triggered stations, $T_{sp,i}$ is the arrival time of the
shower plane to station $i$ and $r_{0,i}$ is the distance of
station $i$ to the shower axis.

The network was trained to output 0 (1)  for a proton (Fe) nucleus
with a training set of 4000 events.

Figure 3 shows the result of applying the trained network to an
independent control sample of 11600 events. Figures 3a,b show the
classification results for protons and Fe respectively. It can
clearly be seen that most of the control events ($80$\% of protons
and $\sim 87$\% iron) are classified correctly.

In order to assess the impact of using information coming from
hybrid events, we performed an additional run including also
$X_{max}$. The corresponding output is shown in figure 4. A
noticeable improvement shows up clearly: $\sim 90$\% of protons
and $\sim 91$\% iron are correctly classified. Furthermore, the
number of ambiguous events with intermediate results between $0$
and $1$ diminishes noticeably producing a cleaner output.

\subsection{Assessing hadronic interaction model dependence}

The same network, trained under the assumption of the validity of
the QGSJet model has been tested below at discriminating showers
described by Sybill hadronic interactions. The results show once
more the stability of the network solution despite the hadronic
interaction model used.

\begin{figure}[t]
\vspace*{2.0mm} 
\includegraphics[width=8.3cm]{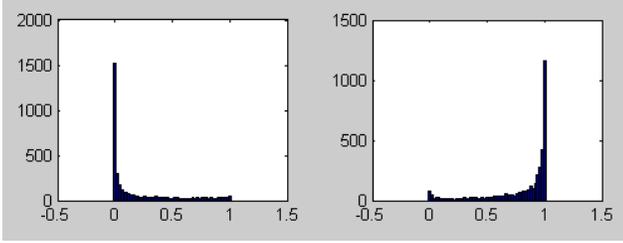} 
\caption{Result of the application of a trained feed-forward
network to an independent control sample of 11600 events triggered
by protons and iron nuclei. The network was trained to output a
value of zero (one) for a proton (iron) primary. Tails, therefore,
correspond to misclassified events. Only surface array information
was included. }
\end{figure}

\begin{figure}[t]
\vspace*{2.0mm} 
\includegraphics[width=8.3cm]{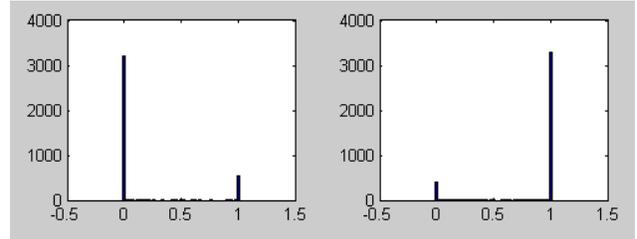} 
\caption{Same as figure 3, but now hybrid events were considered
(basically through the inclusion of X$_{max}$. A much clearer
separation is obtained, despite some events are still
misclassified.}
\end{figure}

\begin{figure}[t]
\vspace*{2.0mm} 
\includegraphics[width=8.3cm]{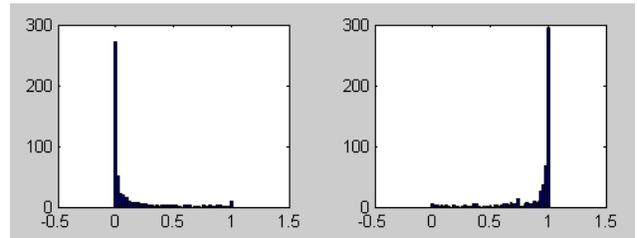} 
\caption{The same neural network of figure 3, trained with EAS
simulations based on the QGSJet hadronic interaction model is used
to discriminate events described by Sybill hadronic interactions.
}
\end{figure}

\begin{acknowledgements}
This work is partially supported by the Brazilian agencies FAPESP
and CNPq .
\end{acknowledgements}

\end{document}